\documentclass[aps,prl,twocolumn,superscriptaddress,groupedaddress,english]{revtex4}
\usepackage[T1]{fontenc}
\usepackage[utf8]{inputenc}
\usepackage{amsmath}
\usepackage{graphicx}
\usepackage{setspace}
\usepackage{babel}
\usepackage{amssymb}
\usepackage{leftidx}
\usepackage[titletoc,toc,title]{appendix}
\usepackage{titlesec}
\usepackage{braket}
\usepackage{bm}
\usepackage{subfig}
\usepackage{xcolor}
\usepackage{caption}
\colorlet{shadecolor2}{yellow!10}
\colorlet{shadecolor}{gray!10}
\usepackage[pdftitle={semi-final},pdfauthor={Wojciech Rzadkowski}]{hyperref}
\hypersetup{colorlinks=flase}
\usepackage{titlesec}
\titleformat{\section}[runin]
  {\normalfont}{\thesection}{1em}{}
\titleformat{\subsection}[runin]
  {\normalfont\bfseries}{\thesubsection}{1em}{} 
\newcommand{\beq}{\begin{equation}}
\newcommand{\eeq}{\end{equation}}
\newcommand{\beqa}{\begin{eqnarray}}
\newcommand{\eeqa}{\end{eqnarray}}

\usepackage{physics}

\begin{document}

\title{Artificial neural network states for non-additive systems}

\author{Wojciech Rzadkowski}
\author{Mikhail Lemeshko}
\affiliation{IST Austria (Institute of Science and Technology Austria), Am Campus 1, 3400 Klosterneuburg, Austria}

\author{Johan H. Mentink}
\affiliation{Radboud University, Institute for Molecules and Materials, Heyendaalseweg 135, 6525 AJ Nijmegen, The Netherlands}
\date{\today}

\begin{abstract}
Methods inspired from machine learning have recently attracted great interest in the computational study of quantum many-particle systems. So far, however, it has proven challenging to deal with microscopic models in which the total number of particles is not conserved. To address this issue, we propose a new variant of neural network states, which we term neural coherent states. Taking the Fr\"ohlich impurity model as a case study, we show that neural coherent states can learn the ground state of non-additive systems very well. In particular, we observe substantial improvement over the standard coherent state estimates in the most challenging intermediate coupling regime. Our approach is generic and does not assume specific details of the system, suggesting wide applications.
\end{abstract}

\maketitle

\section{\emph{Introduction.}}
\label{sec:introduction}
Integration of ideas originating from machine learning into the study of quantum physics has recently attracted great interest, owing to the new possibilities it offers to tackle challenging problems in quantum physics~\cite{biamonte, mlphysical, newtrends}. As pioneered by Carleo and Troyer~\cite{CarleoTroyer}, a particularly appealing approach is to represent the quantum many body wave function by an artificial neural network. This was first demonstrated to quantum spin systems in one and two dimensions and subsequently generalized to bosonic~\cite{FeedForwardBH, SaitoBosons} and fermionic~\cite{RBMStronglyCorrelated, cai2018approximating, choo2020fermionic} systems. Moreover, beyond pure quantum states, artificial neural networks can also accurately represent mixed quantum states in open systems~\cite{PhysRevLett.120.240503, PhysRevLett.122.250502, PhysRevB.99.214306, PhysRevLett.122.250503} and quantum systems at finite temperature~\cite{PhysRevResearch.2.013284}.

However, all these examples involve additive many-body systems, for which by definition the total number of particles is conserved. On the other hand, there is an important class of physical systems which does not satisfy particle number conservation. Besides elementary examples including the hole theory in relativistic quantum theory and solid state physics, as well as the approximate descriptions of superfluidity and superconductivity, this comprises the important class of inherently non-additive quantum impurity systems. Such systems include the paradigmatic Holstein~\cite{holstein1959studies}, Fr\"ohlich~\cite{frohlich1954electrons}, and Su-Schrieffer-Heeger (SSH)~\cite{su1979solitons, su1980soliton} models for an electron(spin) interacting with phonons, the Dicke model~\cite{HEPP1973360, PhysRev.93.99} in quantum optics and the Anderson impurity model~\cite{PhysRev.124.41} for magnetic impurities in metals. Owing to their simplicity, these models play a crucial role in the understanding of quantum many-body effects. Nevertheless, analytic solutions are often available only in limiting cases and to address the full complexity of the problem numerical methods are desired. 

Neural-network quantum states are, in principle, a prospective candidate for efficient representation of such complex quantum many-body states. For example, the restricted Boltzmann Machine (RBM), one of the simplest and most widely used architectures~\cite{melko2019restricted}, exhibits volume law entanglement and can represent even models with long-range interactions~\cite{EntanglementNQS}. Analogous to variational Monte Carlo (VMC) approaches to the Holstein and SSH model~\cite{ohgoe2014variational, karakuzu,PhysRevB.102.125149}, recently the electron-phonon correlation factor was represented using an RBM, while keeping a Jastrow correlation factor for the electron subsystem. Lattice polarons have also been tackled with Gaussian process regression capable of extrapolating across their phase transitions~\cite{PhysRevLett.121.255702}. In addition, the Anderson impurity model has been addressed with machine learning methods to find the Green function~\cite{anderson} and to derive its low-energy effective model~\cite{PhysRevB.101.241105}. However, so far no neural network states exists which directly provide an unbiased estimate of the full many-body wave function of non-additive systems. 

In this Letter we show that efficient neural network states for non-additive systems can be constructed as a feed-forward neural network with outputs inspired from the coherent states well-known from quantum optics~\cite{PhysRev.131.2766}. To investigate the efficiency of this architecture, we consider the Fr\"ohlich model featuring long-range interactions between the phonon degrees of freedom and benchmark it against exact diagonalization. In all cases studied, we find that this approach outperforms the standard mean-field coherent state solution, in particular when impurity-induced phonon-phonon correlations are strong.

\begin{figure}[h]
\centering
\includegraphics[width=\columnwidth]{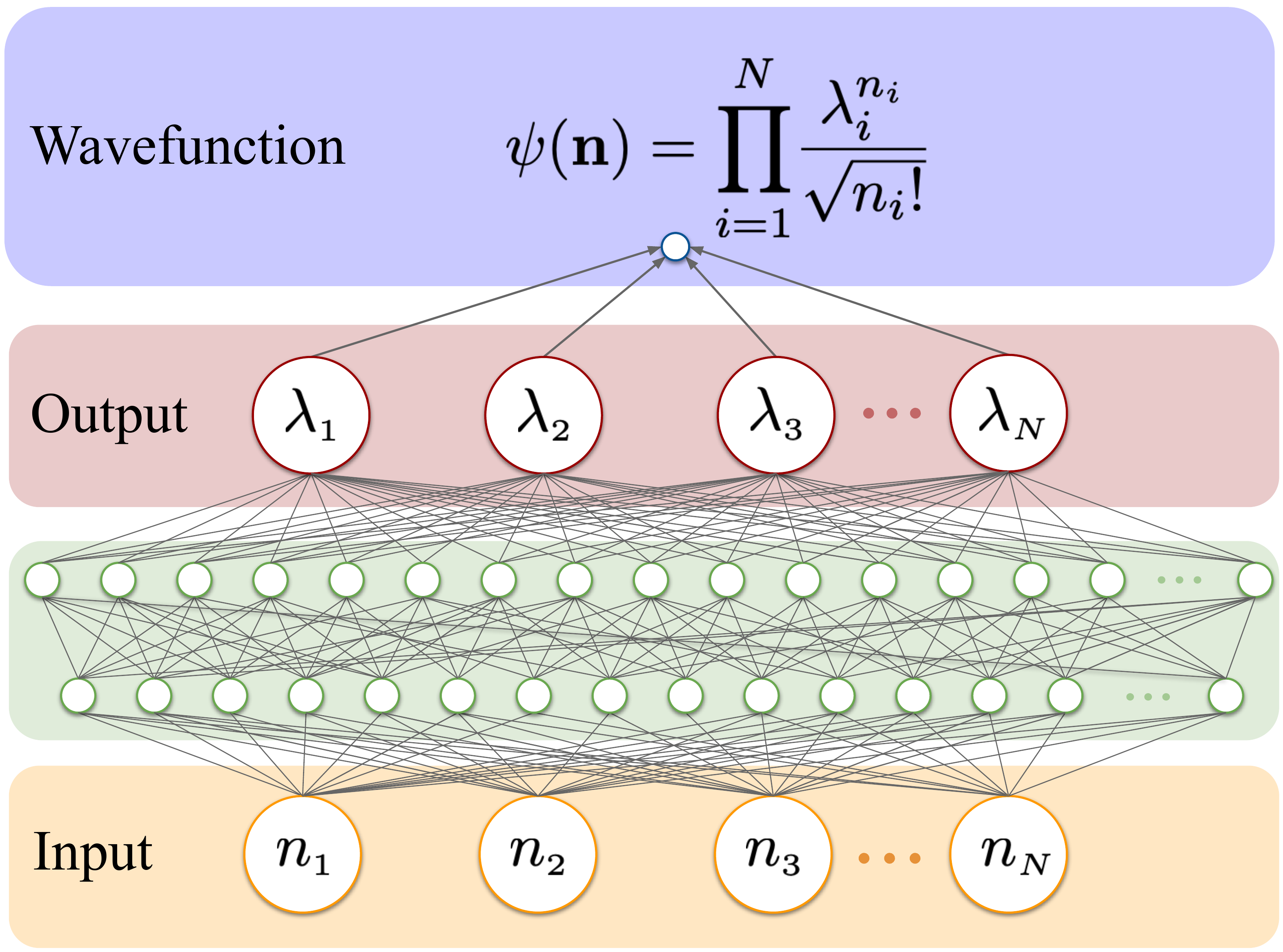}
\caption{Visualization of the NCS approach. The input consists of Fock occupation numbers $n_i$ for each of bosonic modes $i=1,\ldots,N$, corresponding to discrete $k$ values $k_i$. The input is fed to a multilayer perceptron with arbitrary number and size of the hidden layers, see text for details. The number of neurons in the output layer is equal to the number of inputs ($k$-points).  Each of the outputs $\lambda_i$, is transformed using the information from the input, $\lambda_i\rightarrow \lambda_i^{n_i}/\sqrt{ n_i!}$. These numbers are multiplied to form the wavefunction $\psi$. All neurons in the hidden layers are densely connected to all neurons in the neighbouring layers; for clarity of the picture not all of them are visualized with grey lines.}
\label{fig:architecture}
\end{figure}

\section{\emph{Neural-network architecture.}}
\label{sec:architecture}
We use a basis corresponding to bosonic occupations of the system with $\textbf n$ denoting a single bosonic configuration of the whole system:
\begin{equation}
\ket{\textbf n} \equiv \ket{n_1,n_2,\ldots,n_{i},\ldots,n_{N}},
\end{equation}
where $N$ is the number of discrete phonon modes considered. In an RBM architecture, $N$ is equal to the number of visible neurons. However, direct application of an RBM to non-additive systems is not efficient. This can easily been seen by writing the neural-network quantum state as $\ket{\psi(\mathbf{n})}=\psi(\hat{\mathbf{n}})\ket{\mathbf{n}}$; we have $[\psi(\hat{\mathbf{n}}),\hat{N}]=0$, $\hat{N}=\sum_i\hat{n}_i$ for a function $\psi(\mathbf{n})\sim \exp(E(\mathbf{\mathbf{n}}))$, $E$ being linear in $\mathbf{n}$. Hence, by construction an RBM operates in a sector of a given total number of particles\footnote{See the Supplemental Material for more information.}.

To bypass this problem, we propose a neural network inspired from coherent states, which may be termed neural coherent states (NCS) and is illustrated in Fig.~\ref{fig:architecture}. Analogous to a standard coherent state, which for a given $n$ returns an output proportional to ${\lambda^{n}}/{\sqrt{{n}!}}$, with $\lambda$ being the parameter representing the coherent state, we construct a (mutilayer) feedforward neural network taking $\textbf n$ as input. For each configuration, $N$ output numbers $\lambda_i$ are generated, which are subsequently transformed according to: $\lambda_i\rightarrow{{\lambda_i}^{n_i}}/{\sqrt{n_i!}}$. Then these numbers are multiplied to form the wavefunction. If the solution is exactly a coherent state $\lambda$ for each mode $i$, the network simply learns that $\lambda_i\equiv\lambda$ regardless of $n_i$. Correlated solutions are represented by perturbing the numbers $\lambda_i$, such that they depend on the input vector $\mathbf n$ in arbitrary way. This yields the variational Ansatz expressed as:
\begin{equation}
\braket{\mathbf n}{\psi} = \psi(\mathbf n) = \psi(n_1, n_2, \ldots, n_N) = \prod\limits_{i=1}^N\frac{\lambda_i^{n_i}}{\sqrt{n_i!}},
\end{equation}
where $\lambda_i$ is the output of a feedforward neural network (multilayer perceptron) with $M$ layers, i.e.:
\begin{equation}
\bm{\lambda} = \{\lambda_1, \lambda_2, \ldots, \lambda_N\} = h_M(h_{M-1}(\ldots h_1(\mathbf n))), 
\end{equation}
with each of the hidden layer transformations $h_j$ acting on output $\mathbf h_{j-1}$ of the previous layer:
\begin{equation}
h_j\left(\mathbf h_{j-1}\right) = \sigma\left(\mathbf W^{j-1}_j \mathbf h_{j-1} + \mathbf b_j\right)
\end{equation}
where $\mathbf W^{j-1}_j$ and $\mathbf b_j$ are weights and bias of $j$-th layer, while $\sigma(x)$ is an activation function. This function, which introduces nonlinearity in the network, can be chosen from a wide range of classes, such as $\tanh$ (the choice made in this work), sigmoid or ReLU~\cite{nair2010rectified}. 

With this Ansatz, we optimize the variational energy
\begin{equation}
\begin{split}
E=&\frac{\bra{\psi} \hat H \ket{\psi}}{\braket{\psi}}=
\frac{\sum\limits_{\mathbf n, \mathbf n'}\bra{\psi}\ket{\mathbf n}\mel**{\mathbf n}{\hat H}{\mathbf n '}\bra{\mathbf n '}\ket{\psi}}{\sum\limits_{\mathbf n}\bra{\psi}\ket{\mathbf n}\bra{\mathbf n}\ket{\psi}}=\\
=&\frac{\sum\limits_{\mathbf n, \mathbf n'}\psi^*(\mathbf n') \hat H_{\mathbf{n' n}} \psi(\mathbf n)}{\sum\limits_{\mathbf n}|\psi(\mathbf n)|^2}
\end{split}
\end{equation}
using Variational Monte Carlo, by sampling the probability distribution given by $\left|\psi(\mathbf n)\right|^2$~\footnote{See the Supplemental Material for details of the optimization, including considerations specific for non-additive systems, as well as the details of the implementation.}. 

\section{\emph{Hamiltonian representation of impurity problems.}}
\label{sec:representation}
To test the efficiency of the NCS for non-additive systems, we focus on the Fr\"ohlich Hamiltonian as given by:
\begin{equation}
\label{eq:hpretransf}
\hat H = 
\frac{p^2}{2m}
+
\sum_{\mathbf k} \hbar \omega_{0} \hat a^\dagger_{\mathbf k} \hat a_{\mathbf{k}}
+
\sum_{\mathbf k}\left(V_{\mathbf k} \hat a_{\mathbf{k}}e^{- i \mathbf k \mathbf r}+
V^*_{\mathbf k} \hat a_{\mathbf{k}}^\dagger e^{ i \mathbf k \mathbf r}
\right),
\end{equation}
where $\mathbf r$ and $\mathbf p$ denote the position and momentum, respectively, of an impurity characterized by mass $m$. The three terms stand for impurity kinetic energy, bosonic bath energy and the impurity-bath interaction, respectively. The summation extends over all possible wavevectors $\mathbf k$.

Solution of this model is more convenient in the impurity frame, which is achieved by the Lee-Low-Pines transformation~\cite{LLP}. In the sector of zero total momentum, the Hamiltonian becomes: 
\begin{equation}
\label{eq:hdim}
\hat H = 
\frac{
\left(-\sum\limits_{\mathbf k}\hbar \mathbf k \hat a^\dagger_{\mathbf k} \hat a_{\mathbf k} \right)^2
}{2m}
+
\sum_{\mathbf k} \hbar \omega_{0} \hat a^\dagger_{\mathbf k} \hat a_{\mathbf{k}}
+
\sum_{\mathbf k}\left(V_{\mathbf k} \hat a_{\mathbf{k}}+
V^*_{\mathbf k} \hat a_{\mathbf{k}}^\dagger
\right).
\end{equation}
The Lee-Low-Pines transformation removes the impurity degrees of freedom from the Hamiltonian. This maps the problem to a pure problem of interactions between the bosonic modes, at the price of introducing effective interactions between the phonon modes, described by the first term of the transformed Hamiltonian. The transformed  impurity Hamiltonian problem is closer to the lattice boson problems such as the Bose-Hubbard problem, studied earlier with different NQS architectures~\cite{FeedForwardBH, saito2017machine}. However, the problem mentioned earlier that the total number of bosons is not conserved, persists.

\section{\emph{Numerical results.}}
\label{sec:numerical}
To make the Hamiltonian more convenient for numerical computation, we measure energy in units of $\hbar \omega_{0}$. Moreover, we discretize the $k$-grid to include $N$ points $k_i$ ranging from $-k_0$ to $k_0$ with step $\Delta k$. This finally puts the Hamiltonian into the following form in one dimension:
\begin{equation}
\label{eq:Hnumerical}
\hat H =  
\frac{
\left(-\sum_{i}\hbar k_i\hat a_{k_i}^\dagger \hat a_{k_i}\right)^2
}{2m}
+
\sum_{ i} \hat a^\dagger_{ k_i} \hat a_{k_i}
+
\sum_{ i}\left(V_{k_i} \hat a_{k_i}+
V^*_{ k_i} \hat a_{k_i}^\dagger \right)
\end{equation}
In the calculations we further restrict the maximum number of bosons at each mode at a value $n_{\text{max}}$, ranging between 3 and 8, depending on the parameter regime.

To benchmark our results, we compare them with two approaches -- exact diagonalization, and the mean field approach~\cite{DevreeseTutorial}, where the ground state $\ket{\psi_{\textnormal{MF}}}$ is a direct product of coherent states, resulting in energy $E_{\textnormal{MF}}$:
\begin{equation}
\ket{\psi_{\textnormal{MF}}} =
\bigotimes\limits_i\ket{-\frac{V_{k_i}}{1+\frac{\hbar ^2k^2}{2m}}}
,
E_{\text{MF}} = -\sum\limits_k \frac{|V_k|^2}{1+\frac{\hbar ^2k^2}{2m}} .
\end{equation}
By such choice of benchmarks, we are able to quantify the correlations expressed with our Ansatz.

\begin{figure}[h]
\centering
\includegraphics[width=\columnwidth]{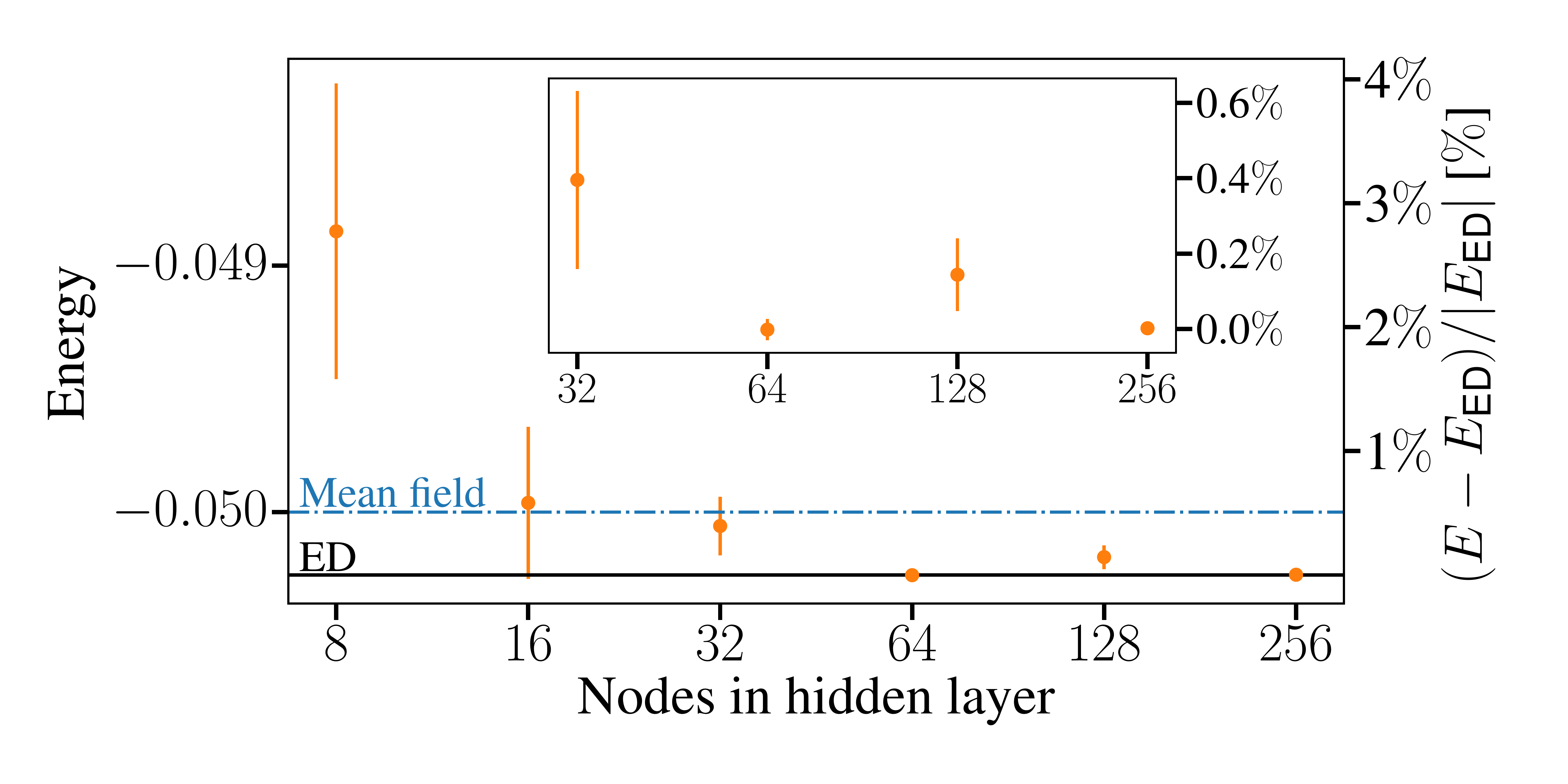}
\caption{
Representative power of the proposed approach. The optimized variational energy (orange dots, in units of $\hbar \omega_0$) is compared with exact diagonalization (ED), see right y-axis for percent scale relative to ED; and mean-field result for a system with 2 bosonic modes. The inset shows percent difference to ED for the four largest network sizes.}
\label{fig:representativepower}
\end{figure}

As the first test, we take a small system with $N=2$ $k$-points $-k_0$ and $k_0$. We fix the impurity-bath potential at $V/(\hbar \omega_0)=0.2$. Moreover, using convenient unit $m_0=\hbar k_0^2/(2\omega_0)$ for the mass, such that $\hbar^2k_0^2 / (2m_0)=\hbar\omega_0$, we fix the inverse mass at   $1/m=0.6\cdot(1/m_0)$. We vary the number of nodes in the single hidden layer, thus changing the number of variational parameters and, consequently, the representative power of the network. For each number of nodes, we optimize the energy and compare the obtained energy with the ED and mean field energy mentioned above. The results are shown in Fig.~\ref{fig:representativepower}.
We observe that with the number of nodes high enough, the variational energy clearly goes below the mean-field one, proving the capability of our approach. For 64 and 256 nodes in the hidden layer, we have obtained an agreement with the ED result within the stochastic error of our variational approach.

\begin{figure}[h]
\flushleft{(a)}
\includegraphics[width=\columnwidth]{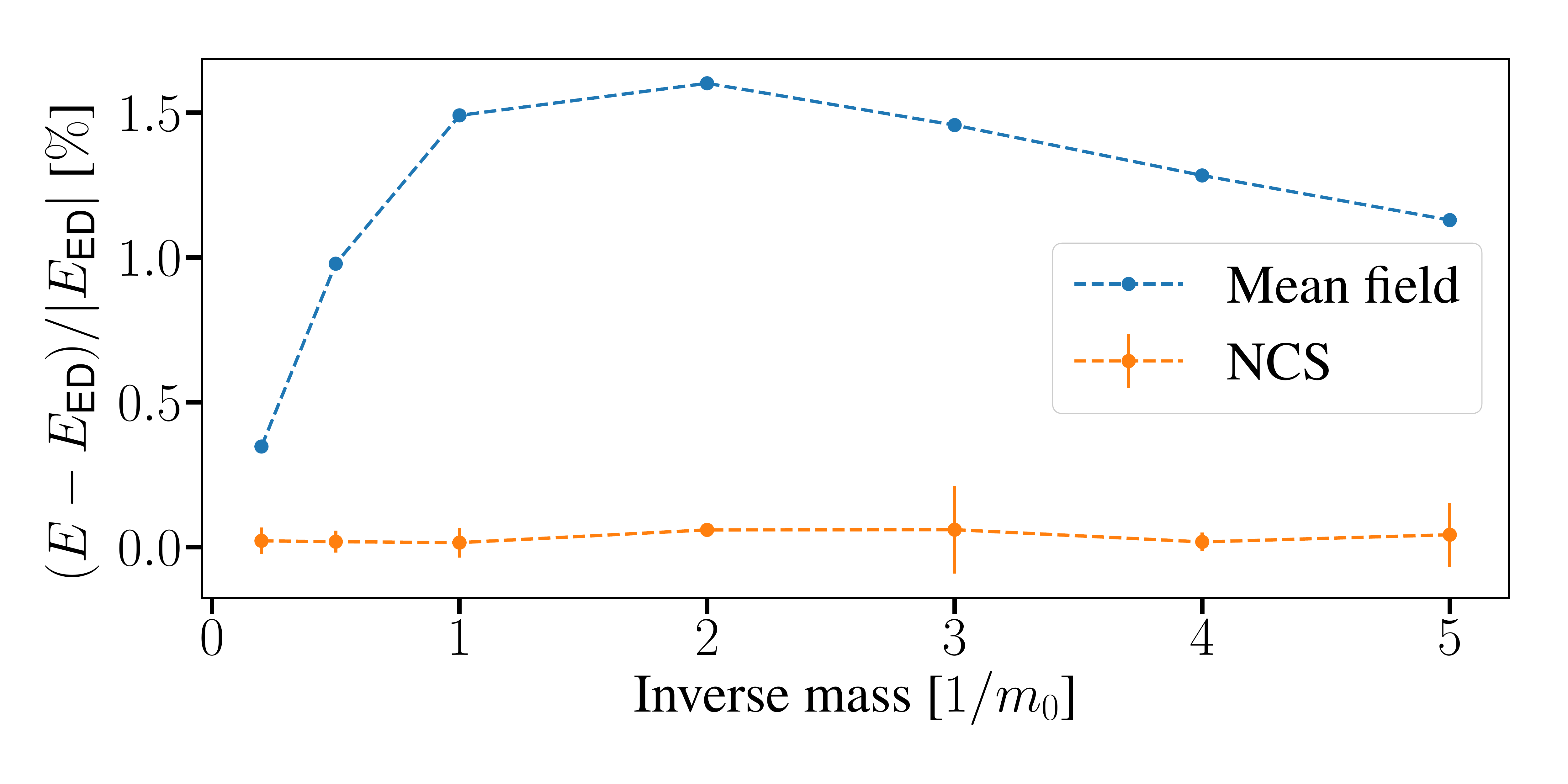}
\flushleft{(b)}
\includegraphics[width=\columnwidth]{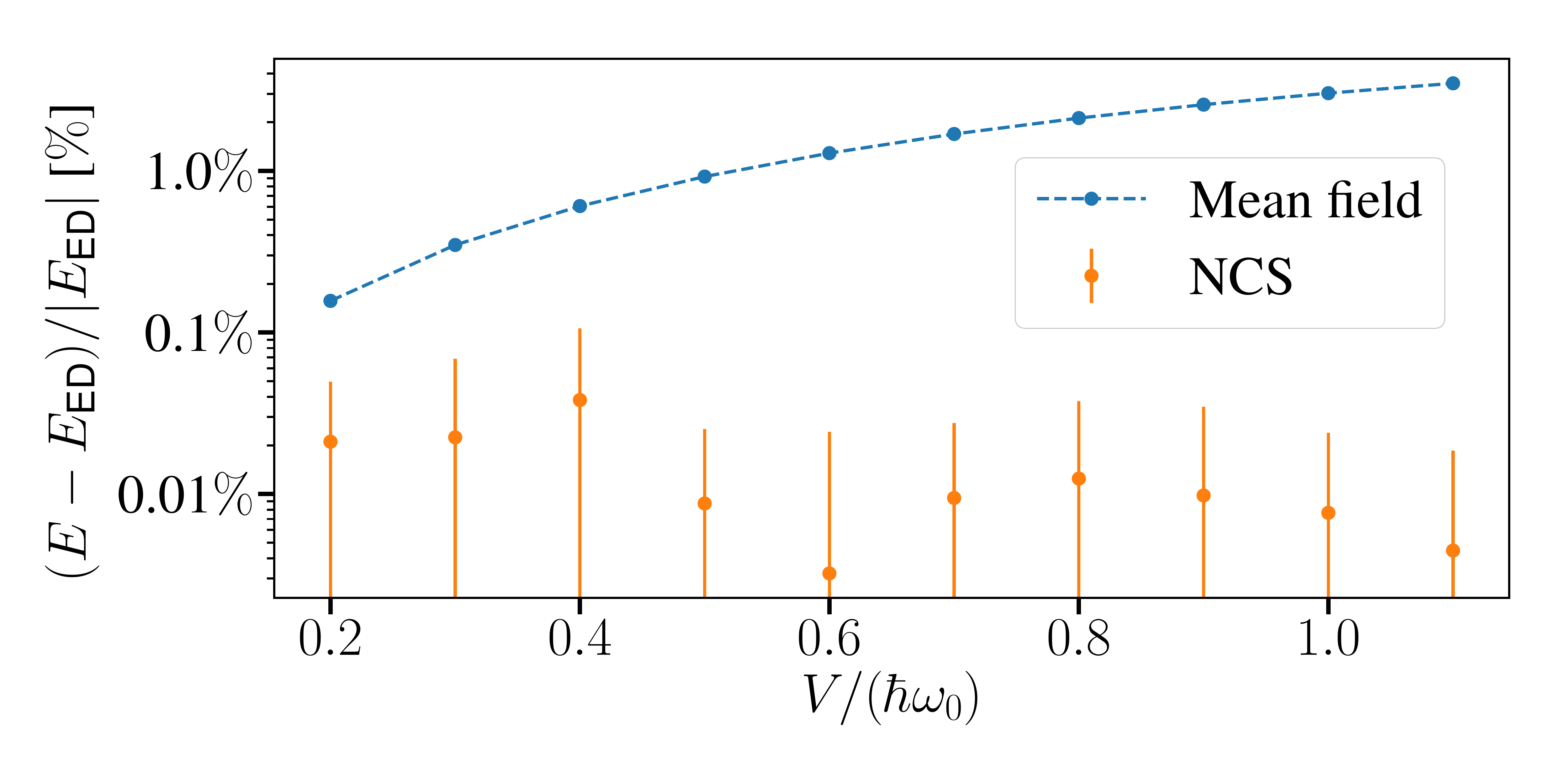}
\caption{The percent difference with respect to ED as a function of: (a) inverse mass in units $1/m_0$, $m_0 = \hbar k_0^2/(2\omega_0)$ (b) impurity-bath coupling $V/\hbar\omega_0$. Our NCS approach (orange) is compared with mean field (blue). Error bars correspond to uncertainty of stochastic estimates of the energy. In panel (b), which features log scale, the error bars stretch infinitely down as the ED result lies within error bars.  The dashed lines guide the eye.} 
\label{fig:mass_V}
\end{figure}

To further evaluate the ability of NCS to express correlations between different bosonic modes, we study the performance of our approach with different impurity mass. Low mass is associated with high correlation level, while high mass brings the Hamiltonian closer to the infinite mass regime, where an analytic solution in the form of coherent state exists. We fix $V$ at $V/(\hbar \omega_0)=0.2$. Then we optimize a network containing 1024 neurons in the hidden layer for different mass values and compare the result with the mean-field approach. The percentage deviation, $100\%\cdot(E-E_{\textnormal{ED}})/|E_{\textnormal{ED}}|$, for both the mean-field and NCS approach is shown in Fig.~\ref{fig:mass_V}(a) as a function of the inverse mass.
Here, we observe very stable performance -- the NCS is able to outperform mean-field approach across the range of (inverse) mass tested. Even at the intermediate regime, where mean-field solution lies $\approx 1.5\%$ above the ED, the NCS is able to accurately express the correlations and agree with ED within the stochastic error. 

Next, we study the performance at different impurity-bath couplings $V$. We fix the inverse mass at $1/m=0.2\cdot(1/m_0)$. Then we optimize the same  network with 1024 neurons in the hidden layer for different values of $V$. The percentage deviation from ED, for both NCS and mean-field approach is shown in Fig.~\ref{fig:mass_V}(b). Here we observe consistently good performance and clear advantage over mean-field across all values of impurity-bath coupling $V>0.1$. Data for $V<0.1$ is not shown, $E_{MF}-E_{ED}<10^{-5}$ and numerical errors in the gradients start to dominate the optimization, making it hard to reach more accurate results. We attribute this with a property of the NCS itself. For such small $V$, the system is very close to the vacuum state. When approaching the coherent state with $|\lambda|=\epsilon\ll1$, leads to a dominance of states $\lambda_i=\epsilon$ for which $\epsilon^0=1$, independent on $\epsilon$. Importantly, our approach easily extends outside the weak-coupling regime, reaching equally accurate results for all $V$, even in the regime $V\sim 1$. 

\begin{figure}[h]
\includegraphics[width=\columnwidth]{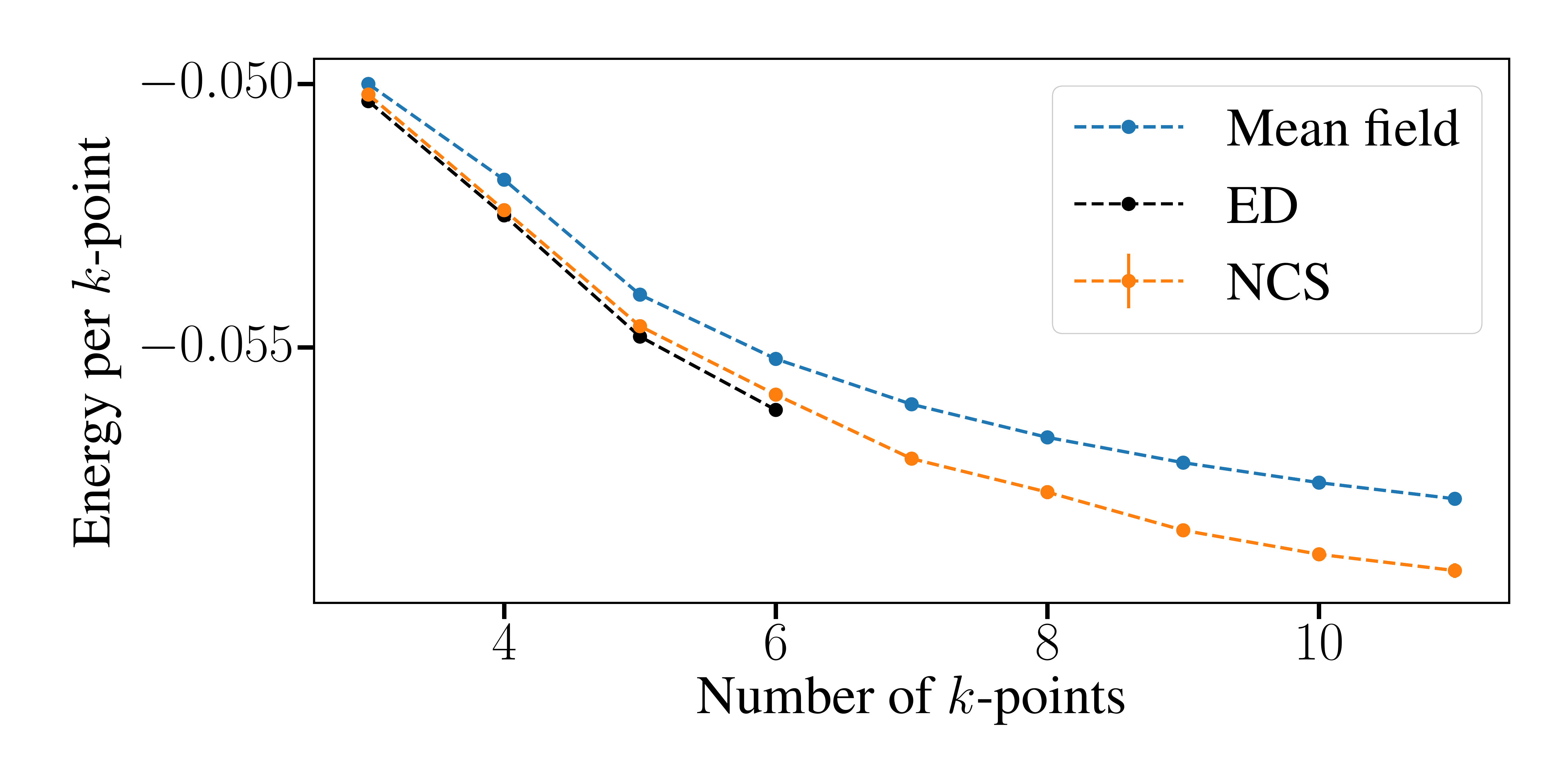}
\caption{The energy divided by number of $k$-points calculated with the NCS approach as a function of the number of $k$-points on an equidistant grid between $k=-1$ and $k=1$. Error bars for NCS approach are smaller than point size. Dashed lines guide the eye.}
\label{fig:continuous}
\end{figure}

So far, all results are obtained for high maximum occupation numbers (up to $n_{\textnormal{max}}=8$) but only a small number of phonon modes. Next we gradually increase the number of $k$-points to benchmark the ability of the NCS approach to express the correlations between a larger number of bosonic modes, beyond a regime where ED is available. To this end we take $k$ being an equidistant grid between $-k_0$ and $k_0$ with a varying number of points. The constant impurity-bath interaction potential is $V_k\equiv0.3$, which corresponds to contact interactions in real space, which is reasonable for one-dimensional systems. We fix the mass at $1/m=2\cdot(1/m_0)$.
Instead of the total energy, we are now interested in energy per number bosonic modes, to avoid a trivial scaling with the number of modes.
In Fig.~\ref{fig:continuous} we show the results of a benchmark against the mean-field approach and, where feasible, exact diagonalization.
We observe that the difference between the energy reached using our NCS state and the mean field solution increases with increasing the number of phonon modes, which is consistent with the fact that the amount of modes that are coupled to each other increases as well. Moreover, within the range where ED is feasible, we observe that the NCS results closely follow ED, while the mean-field energy is systematically higher. The number of network parameters in the single hidden layer network with 1024 hidden neurons used for training this system is much smaller ($\sim$2000 times smaller for 11 $k$-points) than the dimension of the Hilbert space, suggesting great potential to exploit this approach beyond the regime accessible with exact diagonalization.

\section{\emph{Conclusions.}}
\label{sec:conclusion}

In summary, we introduced a new approach to solve non-additive systems with artificial neural networks. By benchmarking against exact diagonalization, we obtained accurate results for small systems and all parameter regimes studied. In particular, we were able to capture the challenging intermediate coupling regime at the same accuracy as weak-coupling results, illustrating that this method provides an unbiased approach to strong correlations in non-additive systems. Natural next steps include benchmarks against other methods for impurity systems and generalizations to other neural network architectures and more complex impurity models,  such as the angulon quasiparticle~\cite{AngulonPRL2015,AngulonPRL2017,AngulonMagnetic} which is the rotational counterpart of the polaron. The main complication is the non-commutative SO(3) algebra describing quantum rotations, which is inherently involved in the angulon problem. Some work in similar direction has already been done for spin models~\cite{vieijra2019restricted}, where irreducible representations of SU(2) were considered as inputs for the network. 
An appealing feature of variational neural network algorithms is their direct extension to unitary quantum dynamics of the system~\cite{CarleoTroyer,  PhysRevB.98.024311, 10.21468/SciPostPhys.7.1.004, PhysRevLett.125.100503, hofmann2021role}. 
This requires generalizing the current approach to complex valued network parameters, yielding the possibility of extension of the presented work to the case of impurity dynamics, understanding of which is a subject of intensive ongoing research~\cite{PhysRevLett.121.026805, PhysRevB.81.085126, PhysRevLett.124.206801, skou2021non, PhysRevLett.110.240601, cherepanov2019far}.

We acknowledge fruitful discussions with Giacomo Bighin, Giammarco Fabiani, Areg Ghazaryan, Christoph Lampert, and Artem Volosniev at various stages of this work. W.R.  is a recipient of a DOC Fellowship of the Austrian Academy of Sciences and has received funding from the EU Horizon 2020 programme under the Marie Sk{\l}odowska-Curie Grant Agreement No. 665385. M. L. acknowledges support
by the European Research Council (ERC) Starting Grant No. 801770 (ANGULON). This work is part of the Shell-NWO/FOM-initiative ``Computational sciences for energy research” of Shell and Chemical Sciences, Earth and Life Sciences, Physical Sciences, FOM and STW.

\end{document}


\title{Supplemental Material for `Artificial neural network states for non-additive systems'}
\author{Wojciech Rzadkowski}
\author{Mikhail Lemeshko}
\affiliation{IST Austria (Institute of Science and Technology Austria), Am Campus 1, 3400 Klosterneuburg, Austria}

\author{Johan H. Mentink}
\affiliation{Radboud University, Institute for Molecules and Materials, Heyendaalseweg 135, 6525 AJ Nijmegen, The Netherlands}
\date{\today}

\maketitle

\setcounter{equation}{0}
\renewcommand{\theequation}{S\arabic{equation}}
\renewcommand\thefigure{S\arabic{figure}}
\section{Non-representability of coherent state by Restricted Boltzmann Machine}
\label{app:nonrepresentability}
In the main text we argued that an RBM operates in a fixed particle number sector. Here we provide an explicit proof that an RBM is not capable of representing the coherent state. This is relevant because the ground state of Fr\"ohlich hamiltonian is coherent in the infinite mass limit and in the mean-field approximation.
Let us, without loss of generality, consider just one $k$-point, i.e. $\mathbf n \equiv n$. The ground state $\ket{-V}$ is coherent and its wavefunction is given by:
\begin{equation}
\label{eq:ourgoal}
\psi_\text{GS}(n)=\braket{n}{\psi_\text{GS}}=
\exp(-\frac{|V|^2}{2})\frac{(-V^*)^n}{\sqrt{n!}}
\end{equation}
It needs to be represented by the RMB variational Ansatz:
\begin{equation}
\label{eq:NQS}
\psi_\text{RBM}(n)=\exp(an)(\exp(b+Wn)+\exp(-b-Wn))
\end{equation}
The task is to find complex numbers $a,b,W$ such that Eq.~(\ref{eq:NQS}) matches Eq.~(\ref{eq:ourgoal}).

This is infeasible. Proof (by contradiction): assume that one can find $a,b,W$ such that $\psi_\text{RBM}(n)=\psi_\text{GS}(n)$. Then equating $\psi_\text{RBM}(n+1)/\psi_\text{RBM}(n)$ with $\psi_\text{GS}(n+1)/\psi_\text{GS}(n)$ we obtain:
\begin{equation}
-\frac{V^\star}{n+1}=e^a\frac{\cosh(b+Wn+W)}{\cosh(b+Wn)}
\end{equation}
The left and right hand sides of this equation have different $n\rightarrow \infty$ limits. The left hand side decreases to 0 with rising $n$ while right hand side tends to 1 with rising $n$. This ends the proof.

\section{Details of the Monte Carlo optimization}
\label{app:optimization}
The derivatives of energy with respect to variational parameters (``forces'') $\mathcal{F}_{\xi}(\mathcal{W}^{(p)})$, are  given by:
\begin{equation}
\label{eq:forces}
\mathcal{F}_{\xi}=\left\langle E_{\text{loc}} O_{\xi}^{*}\right\rangle-\left\langle E_{\text{loc}}\right\rangle\left\langle O_{\xi}^{*}\right\rangle,\quad
E_{\text{loc}}(\mathbf n)=\frac{\sum\limits_{\mathbf n'}\hat H_{\mathbf n' \mathbf n}\psi(\mathbf n')}{\psi(\mathbf n)}.
\end{equation}
Here $O_{\xi}$ are the logarithmic derivatives of the wavefunction with respect to variational parameters, which are indexed by a collective index $\xi$ running across all weights and biases of the model:
\begin{equation}
O_{\xi} = \frac{1}{\psi(\mathbf n)} \frac{\partial\psi(\mathbf n)}{\partial \xi} = \frac{\partial\log\left( \psi(\mathbf n)\right)}{\partial \xi}.
\end{equation}
The quantity $E_{\text{loc}}(\mathbf n)$ is commonly referred to as local energy of the state $\mathbf n$. If the Hamiltonian matrix is sparse, local energy can  be efficiently computed numerically. The gradients calculated using Eq.~(\ref{eq:forces}) can be an input to any optimization method, either inspired by physics, like stochastic reconfiguration~\cite{sorella} and linear method~\cite{linear} or an algorithm chosen from a rich field of approaches originating from the field of machine learning. In this work, we choose the Adam algorithm~\cite{adam}.

The braces $\langle\cdot\rangle$ refer to weighted averages over the probability distribution given by the wavefunction; the average of any quantity $x(\mathbf n)$ is given by:
\begin{equation}
\langle x(\mathbf n)\rangle = \frac{
\sum\limits_{\mathbf n}|\psi(\mathbf n)|^2 x(\mathbf n)}{\sum\limits_{\mathbf n}|\psi(\mathbf n)|^2}.
\end{equation}

Such averages are not tractable as the sum runs over an exponential number of all possible states of the system. One has to resort to stochastic techniques. To sample the system for a given set of variational parameters and get the estimates $\langle \cdot \rangle$ of the logarithmic derivatives and local energies, we use the Metropolis-Hastings algorithm. At first, we choose a random initial configuration $\textbf n^{(0)}$. Here, we choose $\textbf n^{(0)}$ by randomly assigning 0 or 1 with probabilities 1/2 to each of the $k$-points $k_i$. This, in average, results in starting with a vector with mean occupation equal to $N/2$, i.e. half of the $k$-modes. We observed that allowing to start from higher mean occupations can lead to unstable sampling, i.e. that instead of reaching the high-probablity regions of the distribution, the sampling reaches states with mean occupation rising in an uncontrolled way. The possibility of such a behavior comes from the already mentioned fact that the non-additive Hamiltonian connects states with different total numbers of bosons, contrary to e.g. the Bose-Hubbard model.

Given a sample $\mathbf n^{(i)}$, finding the next sample $\mathbf n^{(i+1)}$ in the Metropolis-Hastings algorithm consists of two steps:
\begin{enumerate}
\item choosing a candidate for $\mathbf n^{(i+1)}$
\item calculating the acceptance probability $a(\mathbf n^{(i+1)}|\mathbf n^{(i)})$ and accepting the candidate $\mathbf n^{(i+1)}$ with that probability. If the candidate is not accepted, we remain in state $\mathbf n^{(i)}$.
\end{enumerate}

For step 1, every choice is in principle possible. However, one needs to take into account the tradeoff between efficient exploration of the state space and still having a large acceptance probability. A choice of state $\mathbf n^{(i+1)}$ that differs too little from $\mathbf n^{(i)}$ might require too many steps to explore the state space, while a choice of state differing too much leads to lower acceptance ratio. The solution that we adopt here is the so-called Hamiltonian sampling, where the candidate $\mathbf n^{(i+1)}$ is chosen with uniform probability  from all states connected by the Hamiltonian to $\mathbf n^{(i)}$ (i.e. having nonzero Hamiltonian matrix element $\hat H_{\mathbf n^{(i)}\mathbf n^{(i+1)}}$). We found other intuitive solutions, such as for example one that can be applied to Bose-Hubbard model -- choosing randomly a pair of $k$-sites and swapping a phonon between them, to work much worse. In particular, this solution conserves the boson number, while our Hamiltonian does not. 

For step 2, the acceptance probability is proportional to the probability ratio of accepted states:
\begin{equation}
\label{eq:acceptanceprop}
a\left(\mathbf n^{(i+1)}|\mathbf n^{(i)}\right) \propto 
\frac{p\left(\mathbf n^{(i+1)}\right)}{p\left(\mathbf n^{(i)}\right)}=
\frac{\left|\psi(\mathbf n^{(i+1)})\right|^2}{\left|\psi(\mathbf n^{(i)})\right|^2}.
\end{equation}
Moreover, as already noticed e.g. for Jastrow-type approaches~\cite{ohgoe2014variational}, the detailed balance condition~\cite{MCtextbook} needs to be fulfilled:
\begin{equation}
\frac{p\left(\mathbf n^{(i+1)}|\mathbf n^{(i)}\right)}
{p\left(\mathbf n^{(i)}|\mathbf n^{(i+1)}\right)} = 
\frac{p\left(\mathbf n^{(i+1)}\right)}{p(\mathbf n^{(i)})}.
\end{equation}
Observing that 
\begin{equation}
p\left(\mathbf n^{(i+1)}|\mathbf n^{(i)}\right)=a\left(\mathbf n^{(i+1)}|\mathbf n^{(i)}\right)g\left(\mathbf n^{(i+1)}|\mathbf n^{(i)}\right),
\end{equation}
where $g(\mathbf n^{(i+1)}|\mathbf n^{(i)})$ is the probability that state $\mathbf n^{(i+1)}$ was proposed given $\mathbf n^{(i)}$, and combining with the equality from Eq.~(\ref{eq:acceptanceprop}) we arrive at:
\begin{equation}
a\left(\mathbf n^{(i+1)}|\mathbf n^{(i)}\right)=\frac{|\psi(\mathbf n^{(i+1)})|^2}{\left|\psi(\mathbf n^{(i)})\right|^2}\frac{g\left(\mathbf n^{(i)}|\mathbf n^{(i+1)}\right)}{g\left(\mathbf n^{(i+1)}|\mathbf n^{(i)}\right)}.
\end{equation}
Our choice of update proposal yields $g(\mathbf n^{(i+1)}|\mathbf n^{(i)}))) \propto 1/\langle \langle\mathbf n^{(i)}\rangle\rangle$ where $\langle\langle\mathbf n\rangle\rangle$ is the number of states connected to state $\mathbf n$ by the Hamiltonian.
Hence, the probability of accepting the sample proposal $\mathbf n^{(i+1)}$ given current sample $\mathbf n^{(i)}$ is given by
\begin{equation}
\label{eq:acceptance}
a\left(\mathbf n^{(i+1)}|\mathbf n^{(i)}\right)=\frac{\left|\psi(\mathbf n^{(i+1)})\right|^2}{\left|\psi(\mathbf n^{(i)})\right|^2}\frac{\langle \langle\mathbf n^{(i)}\rangle\rangle}{\langle\langle\mathbf n^{(i+1)}\rangle\rangle}
\end{equation}
The need for the term $\langle \langle\mathbf n^{(i)}\rangle\rangle/\langle\langle\mathbf n^{(i+1)}\rangle\rangle$ again results from the non-conservation of the total number of bosons and is absent (i.e. equal to 1) in systems such as the Bose-Hubbard model or Heisenberg spin.
To improve the sampling stability, we run several Monte Carlo chains in parallel. Samples from each of them are collated to form the set of samples used at a given step of the optimization.

\section{Dealing with occasional Monte Carlo sampler instabilities}
\label{app:sampler}
\begin{figure}[h]
\hspace{-4cm}(a)\newline
\includegraphics[width=0.5\columnwidth]{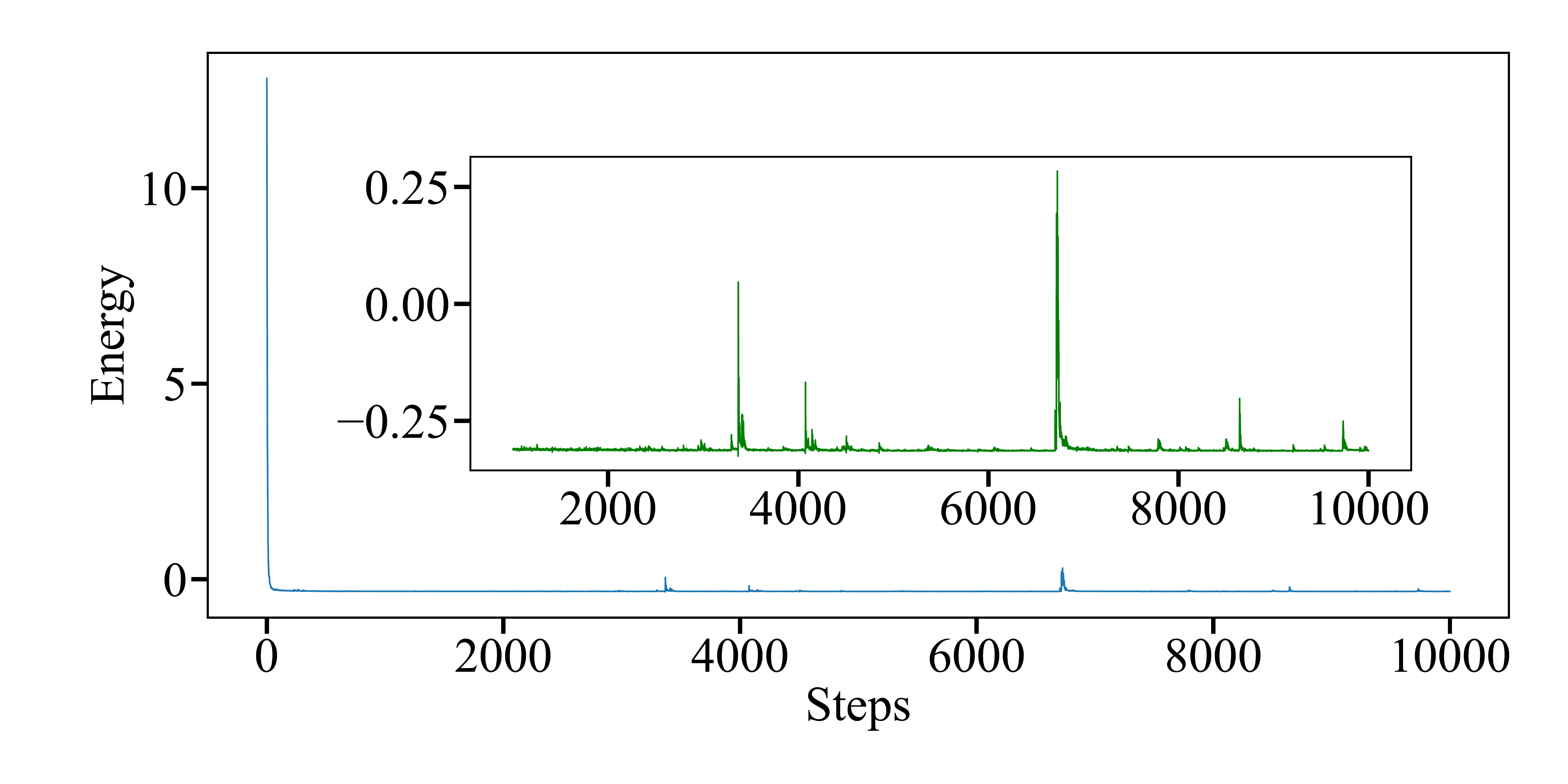}

\hspace{-4cm}(b)\newline
\includegraphics[width=0.5\columnwidth]{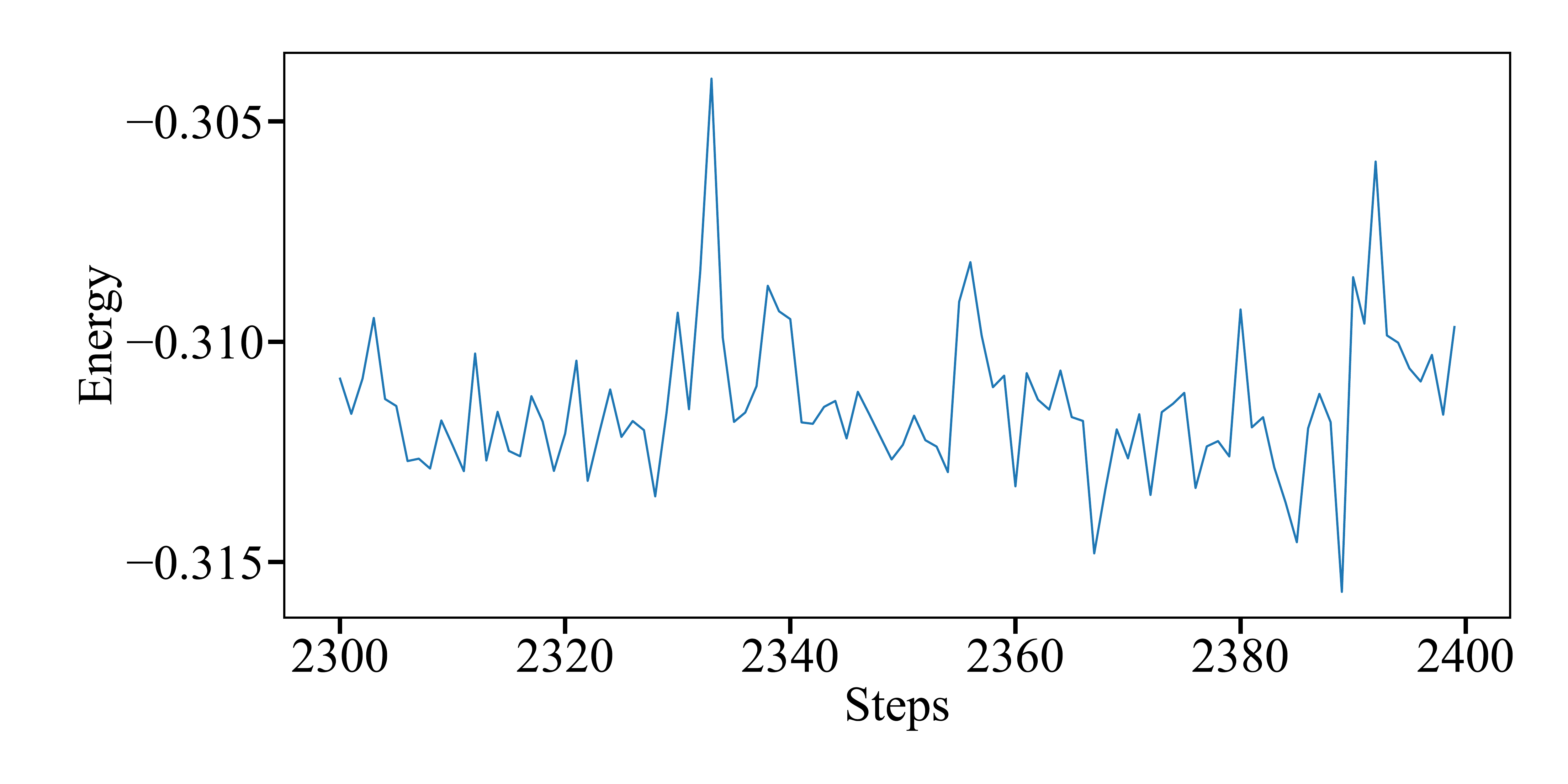}
\caption{Typical example of occasional sampling failures. Panel (a) shows entire 10 000 training steps starting from random parameters, with inset omitting first 1000 steps. Panel (b) shows exemplary 100 continuous training steps from a stable part of the optimization, free of instabilities.}
\label{fig:stability}
\end{figure}

During the training, we have observed that some values of energy are outliers, i.e. they lie off the trend set by preceding and following energies. The phenomenon is illustrated in Fig.~\ref{fig:stability}. We attribute this to a failure of a Monte Carlo chain, i.e. exploring not the entire distribution, but only a region of it. We go around this by always choosing a number (e.g. 100) of subsequent optimization steps that does not contain an energy outlier. Then, we choose the mean from these subsequent steps as the final energy and the standard deviation over these steps as the error of the final energy. Designing a method that excludes such points at runtime, for example by analyzing the discrepancy of energies obtained from the different Monte Carlo chains running in parallel, could be a valuable extension of our work. We have also observed that avoiding very low $V$ regime, increasing the number of hidden nodes and decreasing the learning rate all contribute to a reduction of the frequency and magnitude of such instabilities.

\section{Computer code}
\label{app:code}
To facilitate reproduction of the results and experimentation with the proposed approach, we make the code available online under the following URL: https://github.com/wrzadkow/ncs.

The code uses Jax~\cite{jax2018github} and Flax~\cite{flax2020github} libraries. The Jax library allows execution on CPUs, GPUs and TPUs without change and boasts a high level of control over random number generation with the so-called Threefry counter pseudorandom number generator~\cite{PRNG}. Moreover, the native vectorization functionality provided in Jax is used to parallelize the Monte Carlo chain.  We use Flax, which is one of Jax's neural network libraries, to implement the multilayer perceptron neural network model and the Adam optimizer. 

The code is divided into modules. The energy module implements the Hamiltonian and local energy calculation. The wavefunction module contains functionality related to the variational Ansatz and its optimization. Finally, Monte Carlo sampling is implemented in the sampler module. Further information about the code can be found in the documentation provided directly in the form of docstrings.